\makeatletter
\@namedef{ver@picins.sty}{9999/99/99}
\makeatother

\documentclass[fleqn]{2023SCGE}
\usepackage{ulem}
\begin{document}

\ensubject{subject}

\ArticleType{Article}
\SpecialTopic{SPECIAL TOPIC: Pulsar Timing Array}
\Year{2023}
\Month{August}
\Vol{66}
\No{1}
\DOI{10.1007/s11433-023-2262-0}
\ArtNo{120403}
\ReceiveDate{Aug 29, 2023}
\AcceptDate{Oct 14, 2023; zwwang@pmo.ac.cn, yzfan@pmo.ac.cn}

\title{The nanohertz stochastic gravitational wave background from cosmic string loops
and the abundant high redshift massive galaxies}{The nanohertz stochastic gravitational-wave background from cosmic string loops and the abundant high redshift massive galaxie}

\author[1]{Ziwei Wang}{{zwwang@pmo.ac.cn}}%
\author[1,2]{Lei Lei}{}
\author[3]{Hao Jiao}{}
\author[1,2]{Lei Feng}{}
\author[1,2]{Yi-Zhong Fan}{{yzfan@pmo.ac.cn}}

\AuthorMark{Ziwei Wang}

\AuthorCitation{Ziwei Wang, Lei Lei, Hao Jiao, Lei Feng and Yi-Zhong Fan}

\address[1]{Key Laboratory of Dark Matter and Space Astronomy, Purple Mountain Observatory, Chinese Academy
of Sciences, Nanjing 210023, China;}
\address[2]{School of Astronomy and Space Science, University of Science and Technology of China, Hefei 230026, China;}
\address[3]{Department of Physics, McGill University, Montreal, QC, H3A 2T8, Canada}


\abstract{Recently, pulsar timing array (PTA) experiments have provided compelling evidence for the existence of the nanohertz stochastic gravitational wave background (SGWB). In this work, we demonstrated that cosmic string loops generated from cosmic global strings offer a viable explanation for the observed nanohertz SGWB data, requiring a cosmic string tension parameter of $\log(G\mu) \sim -12$ and a loop number density of $\log N \sim 4$. Additionally, we revisited the impact of cosmic string loops on the abundance of massive galaxies at high redshifts. However, our analysis revealed challenges in identifying a consistent parameter space that can concurrently explain both the SGWB data and observations from the James Webb Space Telescope. This indicates the necessity for either extending the existing model employed in this research or acknowledging distinct physical origins for these two phenomena.}



\maketitle


\begin{multicols}{2}
\section{Introduction}
Since 2020, some pulsar timing array (PTA) experiments, such as the North American Nanohertz Observatory for Gravitational Waves (NANOGrav)~\cite{NANOGrav:2020bcs}, Parkes Pulsar Timing Array (PPTA)~\cite{Kerr:2020qdo,Goncharov:2021oub}, and European Pulsar Timing Array (EPTA) \cite{Chen:2021rqp}, have found evidence for a common spectrum noise at frequencies around $10^{-8}$ Hz, as expected in stochastic gravitational wave (GW) background models \cite{Burke-Spolaor:2018bvk}. Although no correlation between pulsars has been observed for such a signal, its stochastic GW nature has been widely speculated in the literature. Fortunately, the recently improved data analysis of quite a few PTA experiments, including the Chinese Pulsar Timing Array, EPTA, NANOGrav, and PPTA, independently provides strong evidence for a spatially correlated Hellings--Downs signature of the GW background \cite{CPTA,EPTA,NANOGrav,PPTA}. Among the natural sources for the nHz GW background are supermassive black hole binaries in the universe \cite{Ashoorioon:2022raz}. However, the inferred strain amplitude $A_{\rm 1yr}$ is a few times higher than predicted by typical massive black hole evolution models \cite{2021arXiv210811671I}.
Among the various new physics models proposed to interpret the PTA nHz signal(s) \cite{NANOGrav:2023hvm, Antoniadis:2023zhi, Madge:2023cak, Samanta:2020cdk, Battista:2021rlh,Ahmadvand:2023lpp,2023arXiv230617205A,Cai:2023dls,Ellis:2023tsl,Zu:2023olm,Li:2023yaj,Shen:2023pan}, one intriguing possibility involves cosmic strings \cite{Burke-Spolaor:2018bvk,2018PhLB..778..392B,Blasi:2020mfx, Samanta:2020cdk, Ellis:2020ena,NANOGrav:2023hvm,Bian:2022tju,EPTA:2023hof}.

The James Webb Space Telescope (JWST) observed several massive galaxies at high redshift $7\sim9$ in its Early Release Observations. Those massive galaxies with a stellar mass higher than $10^{10}\, M_{\odot}$ were identified using two spectral energy distribution breaks in JWST multiband photometric data \cite{2016Natur.529..502L}. A high star formation efficiency of $\sim 100\%$ is needed to interpret the formation of massive galaxies in the high-redshift universe, 
which is in tension with the $\Lambda$-cold dark matter ($\Lambda \rm CDM$) model \cite{2007Natur.445..183M,Wang:2022jvx}. Moreover, the age of the universe measured is also a concern, which is larger than $\Lambda \rm CDM$ using galaxy scale observed by JWST \cite{2023MNRAS.524.3385G,Melia:2023dsy}. The high-redshift galaxy formation puzzle may be solved by theories like modified dark energy \cite{Wang:2023ros,2023arXiv230712763A,Santini:2022bib}, axion dark matter \cite{2023arXiv230710302B}, and primordial black holes (PBH) \cite{Yuan:2023bvh, Su:2023jno,Huang:2023chx}, among others \cite{Lin:2023ewc,Domenech:2023afs,Zhitnitsky:2023znn,Hutsi:2022fzw,Gong:2022qjx}. The cosmic string loop theory has also been adopted to explain JWST data because cosmic strings can enhance structure formation in the universe \cite{Jiao:2023wcn}.

Cosmic strings are vortex-like topological defects formed during the symmetry-breaking phase transition with a nontrivial first homotopy group of the vacuum manifold \cite{Vilenkin:1984ib, Hindmarsh:1994re,Vilenkin:2000jqa}. Many particle physics theories beyond the Standard Model predict the existence of cosmic strings. For instance, global strings such as axion strings are associated with a spontaneously broken axial $U(1)$ symmetry \cite{PhysRevLett.48.1867, LAZARIDES198221}. Cosmic strings with current and charge are from gauge symmetry breaking, such as superconducting strings carrying electromagnetic current \cite{Witten:1984eb}. In these theories, a network of long strings (with a curvature radius greater than the Hubble scale) and cosmic string loops are inevitably generated in the expanding universe.

The gravitational properties of cosmic strings are characterized by string tension, denoted as $\mu$, which represents the mass of strings per unit length. The string tension $\mu$ is closely related to the energy scale of the phase transition $\eta$, with a relationship of $\mu\sim\eta^2$. In the literature, the string tension $\mu$ is often expressed in terms of the dimensionless parameter $G\mu$, where $G$ denotes the gravitational constant. Determining the value of $G\mu$ through cosmological observations is crucial. By analyzing the angular power spectrum of the cosmic microwave background (CMB) temperature distribution, a stringent constraint of $G\mu<10^{-7}$ has been established \cite{2016PhRvD..93l3503C}. Notably, this robust bound does not rely on any specific assumptions regarding the distribution of string loops. Besides, superconducting cosmic strings are a possible exotic source of electromagnetic signals, like fast radio bursts (FRBs) \cite{Vachaspati:2008su,Cai:2011bi,Cai:2012zd,Yu:2014gea}. FRBs are used to constrain the superconducting cosmic string tension $G\mu \leq 10^{-12}$ in previous works \cite{Brandenberger:2017uwo,Ye:2017lqn,Imtiaz:2020igv}.

Cosmic string parameters can also be constrained by stochastic GW background (SGWB) detected by PTA. Moving or oscillating relativistically, long cosmic strings and loops emit GWs. However, the SGWB from long strings subdominates near nHz frequencies, leading to constraints on string tension looser than what we get from the CMB anisotropy angular power spectrum \cite{CamargoNevesdaCunha:2022mvg}. Thus, in this letter, we focused on gravitational radiation from a scaling distribution of cosmic string loops from global cosmic strings\footnote{Cosmic superconducting strings can also predict GWs from loop decay. However, the electromagnetic radiation decaying channel dominates over the gravitational one when loops shrink to certain sizes, which predicts slightly different number density solutions of the cosmic string loops and GW spectrum at a high-frequency regime\cite{Rybak:2022sbo}.} \cite{Vilenkin:2000jqa,Damour:2000wa,Polchinski:2007rg}.
We present the stochastic GW spectrum from cosmic string loops and use data from PTA nHz signals to constrain the string tension $G\mu$ and the loop number density parameter $N$.

The presence of cosmic strings leads to the confinement of energy, resulting in the generation of overdensities and influencing the process of structure formation. This effect is particularly pronounced in the case of nonlinear density fluctuations, even during the early stages of the universe \cite{Silk:1984xk,Brandenberger:1993by,Vilenkin:2000jqa,Durrer:2001cg,Shlaer:2012rj}. As cosmic string loops accrete matter starting from the time of matter--radiation equality $t_{\rm eq}$, they give rise to highly non-Gaussian fluctuations at very early times. These fluctuations can serve as seeds for the formation of supermassive black holes \cite{Bramberger:2015kua, Cyr:2022urs} and intermediate black holes\cite{Brandenberger:2021zvn}.

In this work, we further investigated the possibility of cosmic string loops contributing to the abundance of stellar halo masses at redshifts of $z = 8$ and $z = 9$, as inferred from the high-redshift galaxies detected from JWST data \cite{Jiao:2023wcn, 2023Natur.616..266L, 2023Natur.Boylan-Kolchin}. This observation potentially poses a challenge to the standard $\Lambda$CDM model. To explore this possibility, we imposed constraints based on recent PTA data.

We organized this letter as follows. In Section II, we provided a brief overview of the scaling distribution of cosmic string loops. In Section III, we calculated the stochastic GW spectrum and utilized it to fit the PTA data. In Section IV, we estimated the stellar mass function of halos seeded by cosmic string loops and compared our findings with observations obtained from the JWST. Additionally, in this section, we compared the parameter space region that can potentially explain the high-redshift galaxies detected by JWST with the region favored by PTA data to assess whether these two observations originate from a common source. Finally, we present our concluding remarks and discussion in the last section. Notably, we adopted natural units throughout this work, specifically setting $\hbar = c = 1$.

\section{Brief Review of the Scaling Solution of Cosmic String Loops}

In the early universe, if there was a breaking of the $U(1)$ symmetry, a network of strings formed and persisted up to the present time. This network comprised long strings with a curvature radius greater than the Hubble scale and cosmic string loops generated through the intersection of these long strings. Because of the much larger curvature radius of the strings compared with their thickness, which is typically of the order $w\sim \eta^{-1}$, the dynamics of cosmic strings can be accurately described by the Nambu--Goto action. This action treats strings as perfect one-dimensional objects. Within this model, cosmic string loops contain cusps and kinks, which refer to points that move with the speed of light and points where discontinuities occur, respectively. It is at these cusps and kinks that the emission of GWs occurs.

The mean separation of long strings is characterized by the correlation length of the scalar field corresponding to the U(1) symmetry breaking. In the expanding universe, the coherence length $\xi$ scales as the causal horizon; that is, $\xi\sim t$ \footnote{Correlations cannot be established faster than the speed of light, so $\xi\leq d_H\sim t$. Meanwhile, if the coherence length is much smaller than the horizon scale, more frequent intersections lead to a rapid energy decay into loops and a quicker increase in $\xi$ compared with $d_H$\cite{Brandenberger:1993by,Vilenkin:2000jqa}.}. This implies that the statistical properties of the long string network remain independent of time as long as all lengths are scaled to the Hubble radius, which is the so-called ``scaling solution.’’ This interesting property of the cosmic string network was studied in the early 1990s \cite{Copeland:1991kz,Austin:1993rg} and verified by numerical simulations \cite{Vanchurin:2005yb}. Furthermore, the scaling solution continues to hold for cosmic string loops because of the competition between their production by intersections of long strings and shrinking by gravitational radiation \cite{Brandenberger:1993by,Vanchurin:2005pa,Ringeval:2005kr,Lorenz:2010sm}.

Specifically, we adopted the one-scale model to describe the density distribution of cosmic string loops \cite{Vilenkin:1981kz,Kibble:1984hp}, wherein all loops with an initial radius $R_i$ are assumed to form at the same time $t_i(R_i)$. Hypothesizing a linear relationship between $R_i$ and $t_i(R_i)$, we proceed as follows:
\begin{equation}
 \frac{R_i}{t_i(R_i)} = \frac{\alpha}{\beta},
 \label{eq:one_scale model}
\end{equation}
where the parameter $\alpha=l/t_i$ represents the ratio of the loop length $l$ to the formation time $t_i$. The value of $\alpha$ must be determined through numerical simulations \cite{BlancoPillado:2011dq,Vanchurin:2005pa,Ringeval:2005kr,Lorenz:2010sm,Blanco-Pillado:2013qja}. In this letter, we chose $\alpha = 0.1$. Additionally, we introduced the parameter $\beta \equiv l/R$, which relates the radius of a loop to its length $l$. The specific value of $\beta$ depends on the particular shape of the cosmic string loop being considered. For this project, we set $\beta = 10$, considering that the shape of the string loops deviated from a perfect circle.

The scaling distribution of cosmic string loops plays a crucial role in the study of the SGWB generated by these loops and in understanding the formation and evolution of structures influenced by them. In the literature, the quantity $n(R,t)$ is commonly used, where $n(R,t)dR$ represents the (comoving) number density of loops at a given time $t$ with a radius ranging from $R$ to $R+dR$. The scaling solution implies that a constant number $N$ of loops are formed per Hubble volume per expansion time, indicating that the comoving number density of loops at the time of their formation is given by
\begin{equation}
 a(t_i)^3n(R_i,t_i) = Nt_i^{-4} = N\alpha^4\beta^{-4}R_i^{-4},
 \label{eq:loop formation}
\end{equation}
Here, $N$ is a parameter determined by the properties of the long string network. Concretely, $N$ is mainly determined by the number of long strings per Hubble volume, as the production of loops is the same as the energy loss of long strings. However, because of the challenges inherent in cosmic string simulations and the significant uncertainties associated with the value of $N$, we treated $N$ as another free parameter that can be constrained through observations in this letter.

As cosmic strings oscillate, they gradually decay by emitting GWs, causing their scales to decrease in the process \cite{PhysRevD.31.3052}. Theoretical frameworks and simulations have indicated that cosmic loops radiate GWs with a constant power given by $\frac{dE}{dt} = -\Gamma G\mu^2 \sim GR^6f^6\mu^2$, where $\Gamma$ represents the gravitational radiation coefficient dependent on the specific dynamics of cosmic strings. To ensure that the energy per length $\mu$ remains unchanged, we assumed that this decay process does not decrease $\mu$. Consequently, the loss of energy leads to a linear decrease in the radius of the cosmic loop, with $\dot R = -\Gamma G\mu\beta$. Thus, a critical radius $R_c(t) = (\Gamma G\mu/\beta) \times t$ can be defined, below which a loop generally cannot maintain its structure for one Hubble time. Meanwhile[Remark: The phrase “on the other hand” is a conjunctive adverb always preceded by “on the one hand” in academic writing. When this phrase is used alone, we typically use other similar phrases or words to convey comparisons and differences or to introduce new topics.], for loops with radii $R \gg R_c$, the back reaction from the slow decay resulting from the emission of gravitational radiation is relatively small.

Given this assumption, we considered that for large loops with $R_i \gg R_c$, their back reaction from gravitational decay is negligible, resulting in $R \simeq R_i$. Meanwhile, for loops with radius $R < R_c$, the number density is approximately independent of radius $R$, specifically $n(R < R_c, t) = n(R_c, t)$, as these loops form within the same Hubble time. We can express the comoving number density of loops per unit radius in the matter era using the following formulation:

\begin{footnotesize}
\begin{equation}
n(R,t>t_{eq}) =
\begin{cases}
    & N\alpha^2\beta^{-2}t_0^{-2}R_i^{-2}, ~~\alpha\beta^{-1}t\leq R <  R_{eq}  \\
    & N\alpha^{5/2}\beta^{-5/2}
        t_{eq}^{1/2}t_0^{-2}R_i^{-5/2},
    ~~R_c \leq R <R_{eq} \\
    & N\alpha^{5/2} \beta^{-5/2}\Gamma^{-5/2}(G\mu)^{-5/2}t_0^{-2}t_{eq}^{1/2} t^{-5/2},
    ~~ R<R_c
\end{cases}
\label{eq:density}
\end{equation}
\end{footnotesize}
where $R_{eq} = \alpha t_{eq}/\beta$ is the radius of loops formed at equality time $t_{eq}$. The first line is for the loops formed in the matter era, while the second and third lines are loops formed in the radiation-dominant epoch. Note that in comoving coordinates, the number density of loops is time-independent when the loop decay is ignored and the time $t$ in the last line comes from the critical radius $R_c(t)$.

Note that we did not consider the impact of loop velocity, but our result is supported by numerical simulations in \cite{Blanco-Pillado:2013qja}. Therein, the distribution of loops has the same radius dependence as in eq.~\eqref{eq:density}, and other parameters are included in the free parameter $N$ in this work. There are other theories on cosmic string loop distribution. A prevalent cosmic string model is the velocity-dependent one-scale model \cite{Martins:1996jp,Martins:2000cs,Martins:2003vd}. The GW signal and corresponding constraint from PTA have been studied previously \cite{NANOGrav:2023hvm}. Besides, SGWB emitted by cosmic string loops based on two different Nambu--Goto string simulations \cite{Blanco-Pillado:2013qja,Ringeval:2005kr} was previously presented as well \cite{Bian:2022tju,LIGOScientific:2021nrg}.


\section{Stochastic Gravitational Waves Production}
In this section, we calculated the stochastic GWs generated by cosmic string loops. We defined the energy spectrum, denoted as $\Omega_{\text{gw}}(f)$, representing the relative energy of GWs within the frequency range from $f$ to $f + d\log f$. To quantify this, we introduced the GW energy per unit frequency and per unit physical volume as $\rho_{\text{gw}(t,f)}$. Dividing this quantity by the critical energy density, we obtained the following relation:
\begin{equation}
\Omega_{\text{gw}}(f) = \frac{\rho_{\text{gw}}(f)}{\rho_c} = \frac{8\pi G}{3H_0^2}f\rho_{\text{gw}}(f),
\label{eq:spectrum_def}
\end{equation}
By carefully evaluating \eqref{eq:spectrum_def} from the contributions of different string loops, the spectrum of the GW was calculated in some previous literature\cite{Ringeval:2017eww,Blanco-Pillado:2013qja}.

When considering cosmic string loops with a given radius $R$, the periodic boundary condition associated with these loops allows for the gravitational emission wave to be decomposed into different harmonics, characterized by frequencies $f_j = \frac{\omega_j}{2\pi} = \frac{j}{\beta R}$. GWs are produced through three distinct types of string interactions: cusp, kink, and kink--kink collision \cite{Damour:2001bk,Polchinski:2006ee,Dubath:2007mf,Polchinski:2007rg}. The power radiated by a cosmic string loop with radius $R$ can be calculated using the following expression:
\begin{equation}
p(R,f) = \sum_{b}N_b\frac{\Gamma_b G\mu^2}{\zeta_b}\sum_{j = 1}^{\infty}j^{-q_b}\delta\left(f - \frac{j}{\beta R}\right)
\label{eq:power single loop},
\end{equation}
The subscript $b$ in this context represents the three different types of microstructures exhibited by string loops in each oscillating mode. In this letter, we only considered cusps dominant in cosmic string loops. For cusps of cosmic string loops, the parameter $\zeta_b$, denoting the Riemann zeta function evaluated at $q_c$, plays a crucial role. Here, $q_c = \frac{4}{3}$ and $\Gamma_b$ represent certain numerical constants that describe the gravitational emission from a single cusp. Specifically, for this model, we have $\Gamma_c = \frac{3(\pi g_c)2}{2{1/3}g^{2/3}}$, where we further utilized the values of $g = \frac{3}{4}$ and $g_c \approx 0.85$ as derived from the model \cite{EPTA:2023hof}.

At the time $t'$, the power per volume emitted by the number density of loops $n(R,t')$ is evaluated after integration over the number density of the cosmic string loops. Considering the redshift from $t'$ to time $t$ for both energy density and frequency itself, the radiation energy density, at time $t$, of the GW is obtained as follows:
\begin{equation}
  \rho_{\text{gw}}(t,f) =\int^t dt' \int  dR \frac{ n(R,t')}{a(t)^3} p\left(R,\frac{a(t)}{a(t')}f\right),
  \label{eq:redshift gravitational wave}
\end{equation}
where we applied the frequency at time $t'$ by $f' =\frac{a(t)}{a(t')}f$. The spectrum today at $t = t_0$ is calculated as follows:
\begin{align}
  \Omega_{\text{gw}}df
  = & \frac{8\pi G^2\mu^2}{3H_0^2}\sum_{j=1}^{\infty}N_c \frac{\Gamma_b j^{-q_c}}{\zeta(q_c)}\int_{0}^{\infty}\frac{dz}{H(z)(1+z)^3}\nonumber \\
  & \times n\left(t(z), \frac{j}{\beta f(1+z)}\right)\frac{j}{2\pi f},
  \label{eq: Stochastic gw from one-scale model}
\end{align}
where we used the relations of $dt = -\frac{dz}{H(z)(1+z)}$ and $z \equiv  \frac{a(t_0)}{a(t)} - 1$. We took the cosmology parameters from Planck \cite{Planck:2018vyg}.

In our analysis, we employed a likelihood method to estimate the parameters $G\mu$ and $N$ in the cosmic string loop model to explain the GW signals observed by PTAs. The likelihood function is given by

\begin{equation}
\mathcal{L}\left(d \vert G\mu, N\right) = \prod_i \frac{1}{\sigma_{i}\sqrt{2\pi}} \exp \left[ -\frac{\left(\Omega_{CSLoop,f}- \Omega_{obs,f}\right)^2}{2(\sigma_{i})^2} \right].
\label{eq:likelihood}
\end{equation}

Here, $\Omega_{obs,f}$ represents the observed relative energy density spectrum at a certain frequency, and $\sigma_{i}$ corresponds to the observational error. Notably, our analysis was conducted in logarithmic space, both for data and errors. For the prior distribution of $G\mu$, we assumed a uniform distribution in logarithmic space, specifically $\log(G\mu) \sim U(-14, -7)$, as guided by constraints from CMB data, which imposed an upper limit of $\log(G\mu) < -7$. Similarly, we set the prior for the parameter $N$ as $\log N \sim U(0, 5)$. Notably, we imposed a strict upper limit on the parameter $N$ because of various physical implications: an excessively large value of $N$ corresponds to a low intercommutation probability of cosmic strings, contradicting the assumptions of scaling solutions in cosmic string theory.

\begin{figure}[H]
\centering
\includegraphics[scale=0.4]{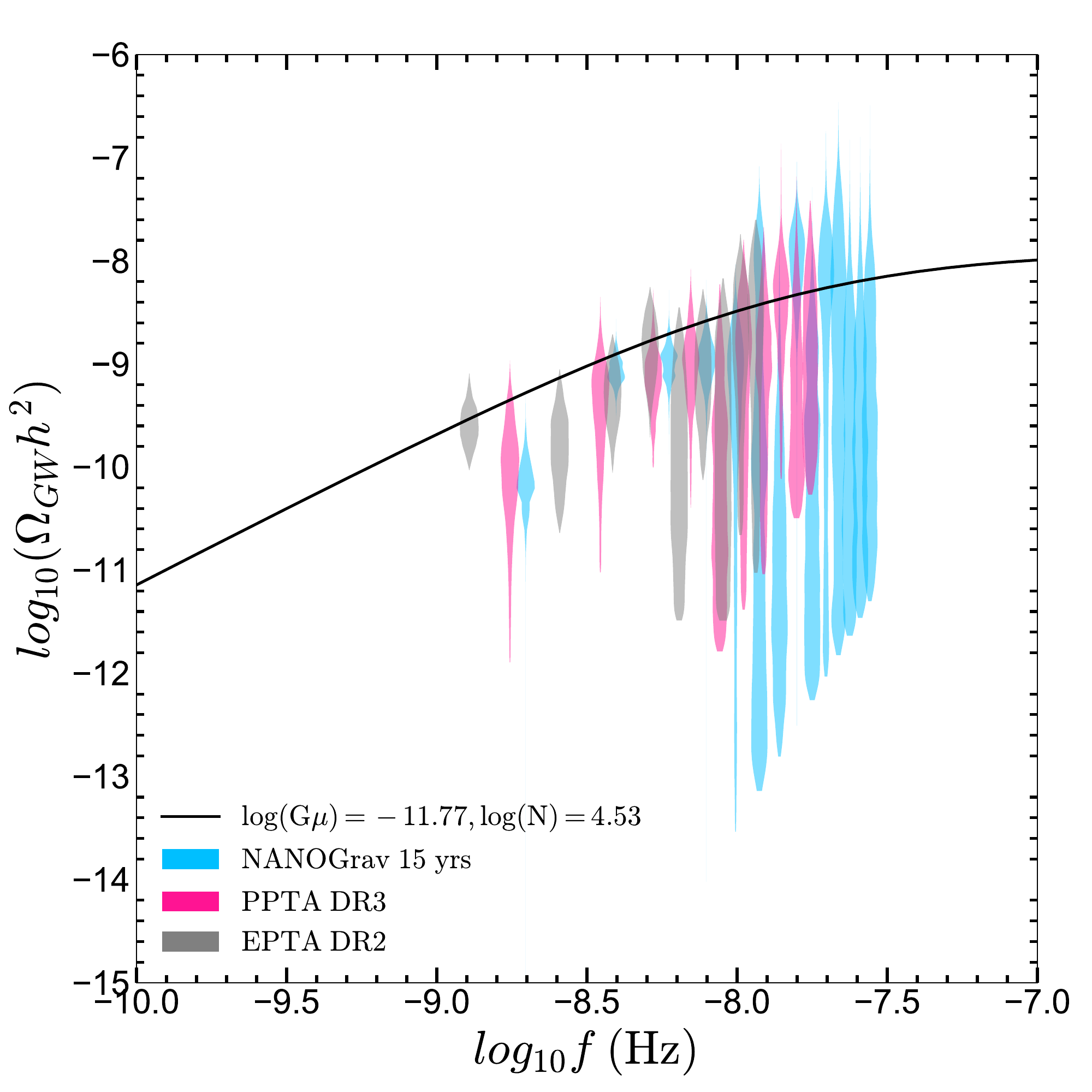}
\caption{GW spectral data of NANOGRAV 15yrs, PPTA DR3, EPTA DR2, and the cosmic string loop GW spectrum model. The sky-blue violin plot shows the NANOGrav data. The gray and black violin plots show the data sets of PPTA and EPTA. The purple solid line shows the cosmic string loop model with parameters $\log (G\mu) = -11.77$ and $\log N = 4.53$
\label{fig:1}}
\end{figure}

\begin{figure}[H]
\centering
\includegraphics[scale=0.5]{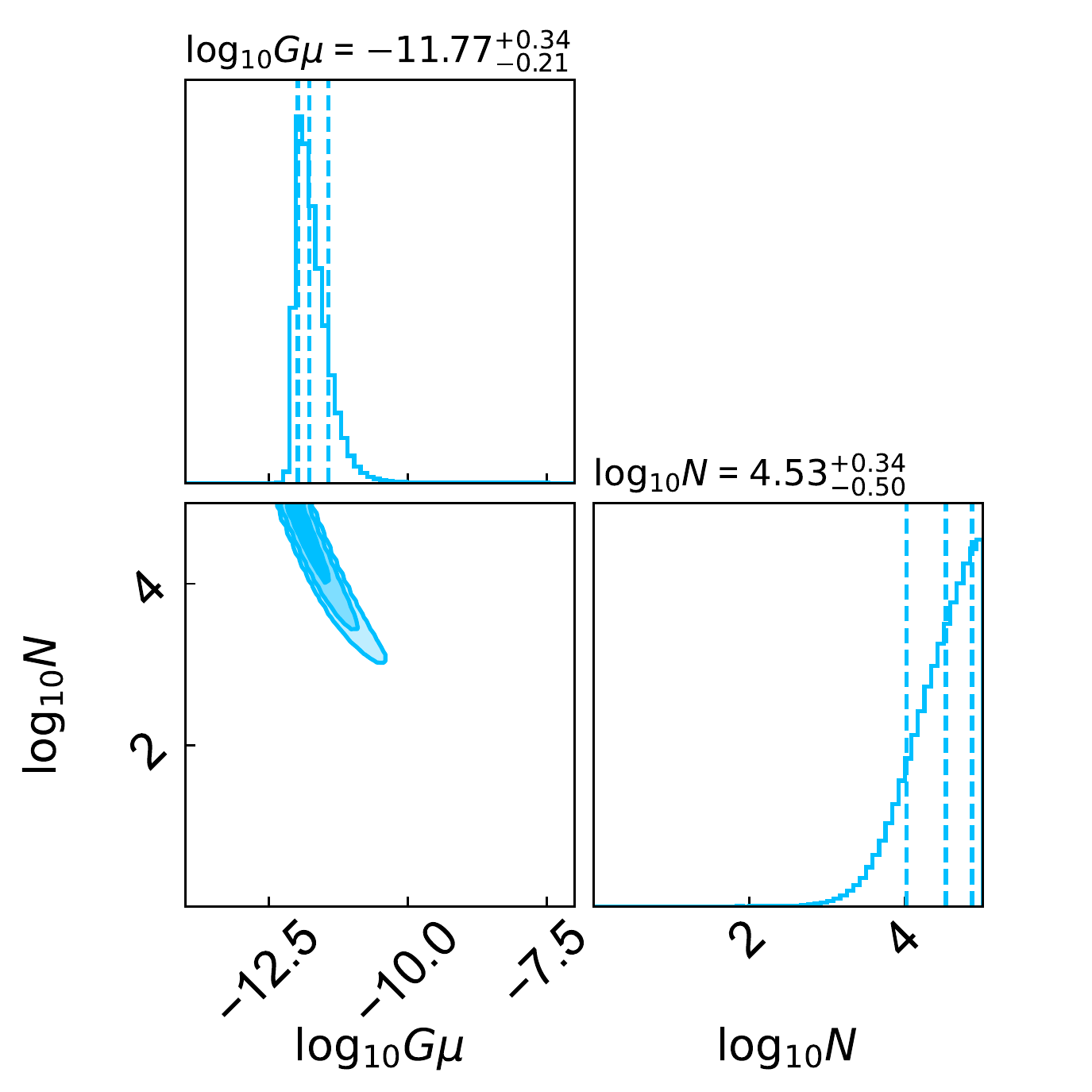}
\caption{
Cosmic string loop model results for fitting the NANOGrav 15 yrs PTA data. The titles on the subplots of the posterior show best-fit values and 1$\sigma$ (68\%) errors. The three coloring contours correspond to the 68\%, 95\%, and 99\% credible regions.
\label{fig:1pdf}}
\end{figure}

Figure~\ref{fig:1pdf} displays the posterior distribution function (PDF) of the cosmic string loop GW model parameters obtained by fitting the NANOGrav 15-year dataset. Notably, the PDF exhibited a region of very low significance or even ruled out in parameter space with larger cosmic string tension $G\mu$. These parameter values corresponded to a flatter GW spectrum shape within the nano-Hz frequency range.  Besides, future observations, such as those made by LISA, notably hold the potential to identify sources contributing to the GW background in different frequency bands\cite{KAGRA:2021kbb}, which may also give more constraints on cosmic string models.

\section{Mass Function From Cosmic String Loops and Early Galaxy Formation}
Unlike Gaussian fluctuations from inflation or its alternative known to provide the dominant contribution to the large structure of the universe, cosmic strings can induce highly non-Gaussian and nonlinear structures at early times. Because the mean distance of cosmic string loops is significantly larger than the turnaround scale during structure formation, we considered the mass accretion onto different loops independently. Suppose mass accretion starts from $t_s$ and lasts till $t$; the accretion mass induced from the string with radius $R$ is given as
\begin{equation}
			\label{eq:mass of accretion for static}
			M = \frac{2G\mu\beta R}{5G}\left(\frac{t}{t_s}\right)^{\frac{2}{3}}.
\end{equation}
We should recognize that mass accretion starts since $t_{eq}$ for loops produced in the radiation-dominant epoch, i.e., $t_s = t_{eq}$, while for loops formed in the matter epoch, they began to accrete matter once they form, yielding $t_s = t_i$.

The mass function of halos sourced by cosmic string loops is derived from the string loop distribution as $dn/dM = dn(R(M),t)/dR \times dR/dM$; that is,

\begin{footnotesize}
\begin{equation}
    \frac{dn}{dM} =
    \begin{cases}
        & \frac{24}{125}N \alpha^4\beta^{-1}t^{-2} \mu^3 M^{-4}, ~~ M > M_{eq}; \\
        & \sqrt{\frac{8}{125}}N\alpha^{5/2}\beta^{-1}t_0^{-2}t^{3/4} t_{eq}^{-1/4}\mu^{3/2} M^{-5/2}, M_{eq} > M > M_{c};  \\
        & \sqrt{10}N\alpha^{5/2}\beta^{-3}G^{-2}\Gamma^{-2}t^{-25/3}t_{eq}^{5/6}\mu^{-5/2} M^{-1/2}, M_c > M
    \end{cases}
\end{equation}
\end{footnotesize}
where $M_{eq}$ is the mass of the halo, which seeds from the cosmic string loops with radius $R_{eq}$ formed exactly at equality time $t_{eq}$, and halos with mass below $M_{c}$ are not expected to accrete sufficiently because of the gravitational decay of cosmic string loops.

The cumulative stellar mass density $\rho (>M_{\star},z)$ is
\begin{equation}
			\label{eq:CSMD}
			\rho (>M_{\star},z) = \int_{M_{\star}}^{\infty}\frac{dn}{dM_{\star}}M_{\star}dM_{\star},
\end{equation}
where the stellar mass $M_{\star}= \epsilon f_b M$ can be estimated from a cosmic string loop-seeded halo mass $M$, the stellar formation efficiency (SFE) $\epsilon=0.2$, and baryon fraction$f_b=0.156$ based on a previous argument \cite{Jiao:2023wcn}. The peak values of SFE
were $\epsilon<0.21$ in high-redshift galaxies \cite{2023arXiv230505679S,2022ApJS..259...20H,2018MNRAS.477.1822M}.
The accretion of baryons in this process has little impact on the linear growth of galaxies and halo masses as the time period is earlier compared with the regular galaxy formation.

\begin{figure}[H]
\centering
\includegraphics[scale=0.3]{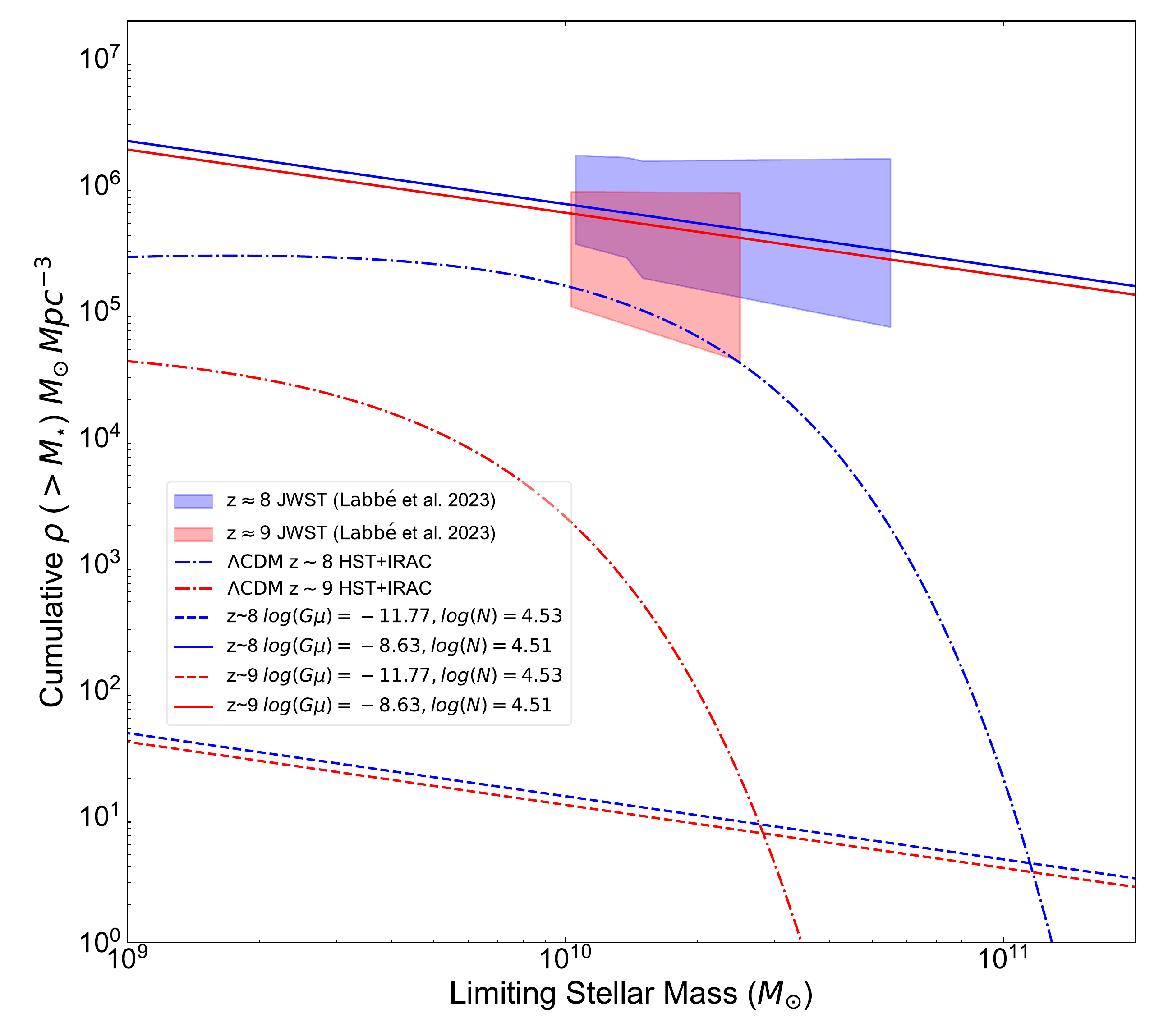}
\caption{Cosmic string loop model fit to the high-redshift JWST massive galaxy stellar mass density data. The cumulative stellar mass density data were taken from a previous study~\cite{2023Natur.616..266L}. Blue and red blocks show JWST data at redshift $z\,[8\sim9]$ from a previous work \cite{2016Natur.529..502L}. The solid lines show a cosmic string loop model with parameters that can explain the JWST data. Dot-dashed lines show the $\Lambda \rm CDM$ stellar mass density model observed by the Hubble Space Telescope and Spitzer Space Telescope Infrared Array Camera. Dashed lines show the cosmic string loop model with PTA best-fit parameters.
\label{fig:2}}
\end{figure}

The cosmic string loop model has been proposed as a potential explanation for the observed excess in stellar mass density among high-redshift galaxies detected by JWST \cite{Jiao:2023wcn}. We used the same model but with cosmic string loop parameters inferred from PTA 15-year data. This allowed us to evaluate the cumulative density $\rho$ and compare it with the stellar mass of massive galaxies from the JWST observations. The corresponding results are presented in Figure \ref{fig:2}.

Upon analysis, cosmic string loops characterized by the parameters of $G\mu \sim \mathcal O(10^{-11} - 10^{-12})$ and $N \sim \mathcal O (10^3 - 10^4)$ evidently predicted a very low stellar mass density, as determined by the standard Press--Schechter formalism, when compared with the observations from the JWST data. A satisfactory explanation of the JWST data by cosmic string loops would require a much larger string tension of $G\mu \sim 10^{-7}$. Figure \ref{fig:2} illustrates the optimal fit of cosmic string loop parameters to the JWST data.

\section{Discussion and Conclusion}
\begin{figure}[H]
\centering
\includegraphics[scale=0.4]{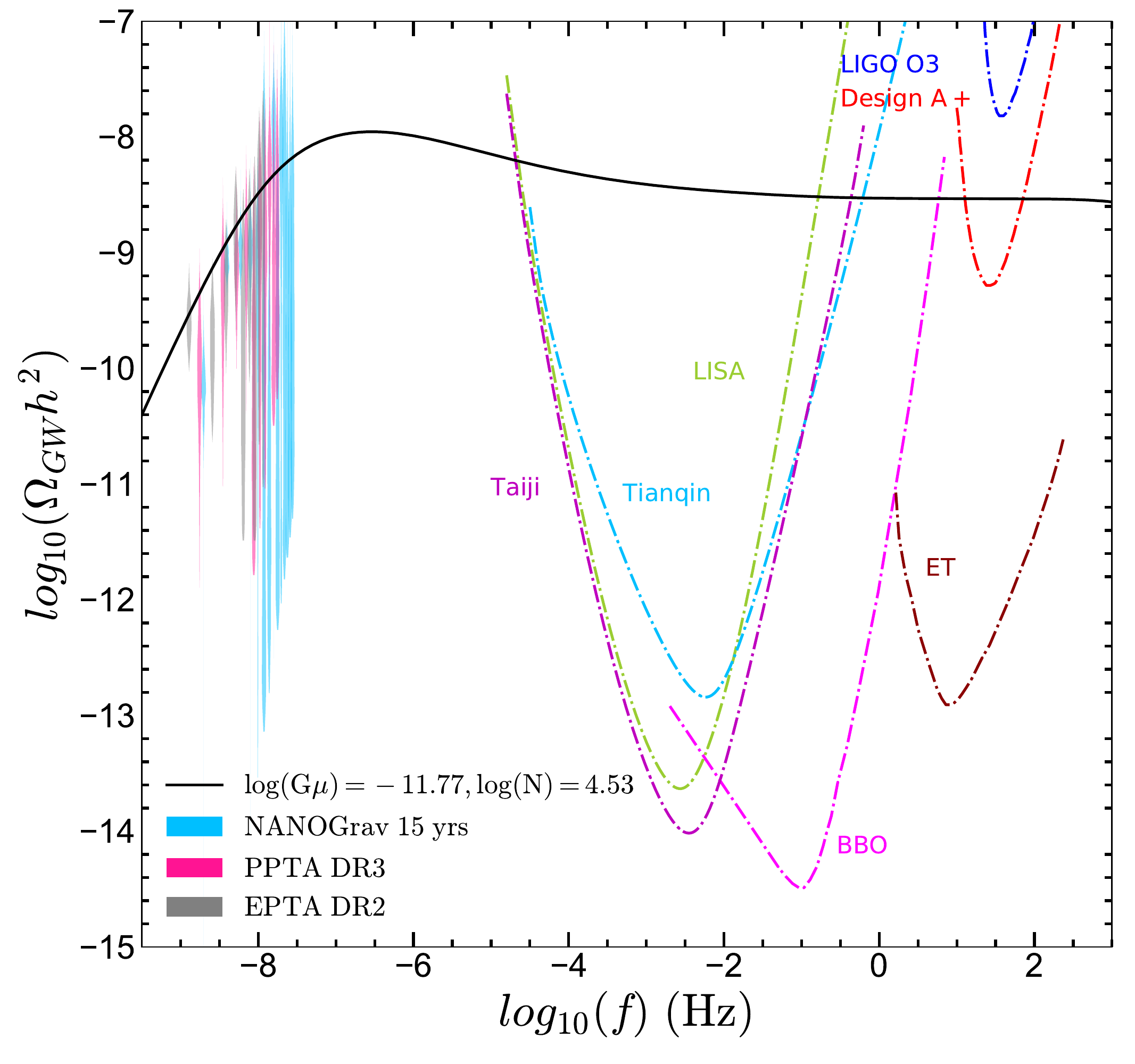}
\caption{GW spectrum data of NANOGRAV 15yrs, PPTA DR3, and EPTA DR2, the cosmic string loop GW spectrum model, and the sensitivity curves of GW observation projects. The sky-blue violin plot shows the NANOGrav data. The gray and black violin plots show the data sets of PPTA and EPTA. The black solid line shows the cosmic string loop model with parameters $\log(G\mu) = -11.77$ and $\log N = 4.53$. The dot-dashed lines in different colors show several GW observation project sensitivity curves, including that of LIGO\cite{LIGOScientific:2014pky, KAGRA:2021kbb}, 
eLISA\cite{Moore:2014lga}, BBO\cite{Crowder:2005nr}, Taiji\cite{10.1093/nsr/nwx116,2020ResPh..1602918L,2021PTEP.2021eA108L}, 
ET\cite{Mentasti:2020yyd,Cho:2022awq}, and Tianqin\cite{TianQin:2015yph,Liang:2021bde}.}
\label{fig:4}
\end{figure}

A groundbreaking advancement in astronomy has been achieved with the discovery of a common spectrum noise at frequencies around $10^{-8}$ Hz, accompanied by a spatially correlated Hellings--Downs signature in the PTA experiments. This remarkable finding strongly suggests the successful detection of nanohertz SGWB. Although massive black hole binaries remain a plausible source, the inferred strain amplitude $A_{\rm 1yr}$ appears to be several times higher than what conventional massive black hole evolution models had predicted. Consequently, this discrepancy indicates the potential necessity for a new and innovative physical model to explain these intriguing observations. Furthermore, the high-frequency plateau of the stochastic GW may be detected by future ground-based or space-based GW experiments \cite{LIGOScientific:2014pky, KAGRA:2021kbb, Crowder:2005nr, Kawamura:2020pcg, LISACosmologyWorkingGroup:2022jok, Crowder:2005nr, 2020ResPh..1602918L, AEDGE:2019nxb, Badurina:2021rgt, Mentasti:2020yyd, Cho:2022awq, Liang:2021bde, Moore:2014lga}. Figure~\ref{fig:4} shows the sensitivity curves of different future GW observation projects \cite{LIGOScientific:2014pky, KAGRA:2021kbb, Crowder:2005nr, Kawamura:2020pcg, LISACosmologyWorkingGroup:2022jok, Crowder:2005nr, 2020ResPh..1602918L, AEDGE:2019nxb, Badurina:2021rgt, Mentasti:2020yyd, Cho:2022awq, Liang:2021bde, Moore:2014lga}. Notably, in this work, we discussed cosmic string loops as soliton solutions of $U(1)$ global symmetry breaking. There are also other types of cosmic strings, such as cosmic superstrings \cite{Ellis:2020ena} and metastable strings \cite{Ahmed:2023pjl, Lazarides:2023ksx}, which can explain the stochastic GW signals from PTA experiments. Additionally, different scenarios are summarized and compared on the basis of their Bayes factors \cite{NANOGrav:2023hvm, Ellis:2023oxs, Bian:2023dnv, Wu:2023hsa}.

The mass functions of several structures in the universe are good tools to test new physics and cosmology models that may impact structure formation differently to explain that either the abundance of PBHs or high-redshift massive galaxies indicate matter perturbation deviating from standard LCDM cosmology \cite{Cai:2023ptf}. Specifically, The JWST excess of high-redshift massive galaxies was explained by several models within new physical processes beyond $\Lambda\rm CDM$ \cite{2023MNRAS.524.3385G, Melia:2023dsy, Wang:2023ros, 2023arXiv230712763A, 2023arXiv230710302B, Santini:2022bib, Yuan:2023bvh, Su:2023jno}. The PBHs from physical process beyond $\rm \Lambda CDM$ can affect the formation of small-scale structures in the observable universe, and the PBH mass function could be tested in JWST observations of supermassive black holes \cite{Cai:2023ptf}. The cosmic string can also be tested by galaxy formation with JWST observations \cite{Jiao:2023wcn}. The large-scale structure observations have provided good model tests for cosmological evolution and the Big Bang. Future observations of small-scale structures and galaxy formation should produce more results on this matter.

In this work, we demonstrated that cosmic string loops, characterized by parameters $G\mu \sim \mathcal O(10^{-11}- 10^{-12})$ and a loop number density $N \sim \mathcal O(10^3 - 10^{4})$, can reasonably account for data on the nanohertz stochastic GW background. However, the same model of cosmic string loops falls short in explaining the abundance of high-redshift massive galaxies recently detected by JWST, as fewer numbers of early galaxies are predicted in the same parameter regime $G\mu$. We propose that either the model employed in this study must be expanded or these two phenomena may have distinct physical origins\cite{Parashari:2023cui}. Besides, the accumulation of more robust data from spectroscopic observations should refine the initial findings from JWST\cite{Wang:2023xmm}. Further investigations of these possibilities are warranted in the near future.

\Acknowledgements{
We thank Robert Brandenberger, Yao-Yu Li, Yi-Ying Wang, Qiang Yuan, Chi Zhang and Lei Zu for useful discussions. We thank Yi-Ming Hu and Zheng-Cheng Liang for providing the sensitivity curves of Tianqin, Taiji and LISA gravitational wave observatories. 
This work is supported by the National Key Research and Development Program of China (No. 2022YFF0503304), the Natural Science Foundation of China (No. 11921003), the New Cornerstone Science Foundation through the XPLORER PRIZE, the Chinese Academy of Sciences, and the Entrepreneurship and Innovation Program of Jiangsu Province. HJ is grateful for hospitality of the Institute for Theoretical Physics, the Institute for Particle Physics and Astrophysics of the ETH Zurich during the completion of this project}

\InterestConflict{The authors declare that they have no conflict of interest.}


\section*{Acknowledgements}
\bibliographystyle{scpma}
\bibliography{biblio.bib}








\end{multicols}
\end{document}